\documentclass[fleqn,usenatbib]{mnras}

\usepackage{newtxtext,newtxmath}
\usepackage{caption}

\usepackage[T1]{fontenc}

\DeclareRobustCommand{\VAN}[3]{#2}
\let\VANthebibliography\thebibliography
\def\thebibliography{\DeclareRobustCommand{\VAN}[3]{##3}\VANthebibliography}

\usepackage{graphicx}	
\usepackage{amsmath}	
\usepackage[dvipsnames,table]{xcolor}
\usepackage{hyperref}
\usepackage{ulem}

\title[\textit{Fermi}-LAT sources redshift prediction with \texttt{CatBoost}]{Redshift prediction of \textit{Fermi}-LAT gamma-ray sources using \texttt{CatBoost} gradient boosting decision trees}

\author[J. Coronado-Blázquez]{
Javier Coronado-Blázquez$^{1}$\thanks{E-mail: j.coronado.blazquez@gmail.com}
\\
$^{1}$Telefónica Tech IoT \& Big Data, Ronda de la Comunicación s/n, Madrid, Spain}

\date{Accepted 2023 March 14. Received 2023 March 13; in original form 2023 January 4}

\pubyear{2023}

\begin{document}
\label{firstpage}
\pagerange{\pageref{firstpage}--\pageref{lastpage}}
\maketitle

\begin{abstract}
The determination of distance is fundamental in astrophysics. Gamma-ray sources are poorly characterized in this sense, as the limited angular resolution and poor photon-count statistics in gamma-ray astronomy makes it difficult to associate them to a multiwavelength object with known redshift. Taking the 1794 active galactic nuclei (AGNs) with known redshift from the \textit{Fermi}-LAT latest AGN catalog, 4LAC--DR3, we employ machine learning techniques to predict the distance of the rest of AGNs based on their spectral and spatial properties. The state-of-the-art \texttt{CatBoost} algorithm reaches an average 0.56 R2 score with 0.46 root-mean-squared error (RMSE), predicting an average redshift value of $z_{avg}=0.63$, with a maximum $z_{max}=1.97$. We use the \texttt{SHAP} explainer package to gain insights into the variables influence on the outcome, and also study the extragalactic bakground light (EBL) implications. In a second part, we use this regression model to predict the redshift of the unassociated sample of the latest LAT point-source catalog, 4FGL--DR3, using the results of a previous paper to determine the possible AGNs within them.
\end{abstract}




\section{Introduction}

The Large Area Telescope on board the NASA \textit{Fermi Gamma-ray Space Telescope} (\textit{Fermi}-LAT, 2008--) continuously monitors the sky searching for gamma-ray sources \cite{fermi_instrument_paper}. \textit{Fermi}-LAT is able to observe the energy band from $\sim$20 MeV to more than 300 GeV, thanks to its pair conversion technology.

The gamma-ray sky presents a rich and complex environment, and may be decomposed in several pieces, such as the isotropic gamma-ray background (IGRB), the interstellar gamma-ray emission, and individual point-like and extended sources. The IGRB and the Galactic diffuse emission represent the main challenge to detect point-sources, due to spectral confusion and photon spill over, given the poor angular resolution of the LAT compared to other wavelengths \cite{2016ApJS..223...26A}. This effect is especially strong at low Galactic latitudes, in the vicinity of the Galactic plane, where the emission of the Milky Way (MW) surpasses any other component\footnote{This diffuse emission is estimated to be responsible for $\sim80\%$ of all LaT-detected photons \cite{Ackermann_2012}.}.

Contrary to cosmic rays (CRs), the direction of propagation of gamma rays remains (almost) unperturbed, which allows for a sky localization. These objects, which are not transient phenomena (although they may experiment flaring periods), are associated with astrophysical sources which can be located in the MW or outside it.

The process of statistical association between a gamma-ray source and a known astronomical object is not trivial, and requires a careful, multiwavelength analysis with several instruments \cite{2017ApJ...838..139S}. The LAT has been able to detect a rich panoply of sources, being active galactic nuclei (AGN) the most frequent.

The current understanding of the AGN landscape is a unified scheme, depending on the radio-band emission and the existence of collimated jets, as well as their orientation towards the Earth. This taxonomy gives rise to objects such as blazars (bll), quasars, or Seyfert galaxies \cite{Urry_1995}. In particular blls --AGNs with a relativistic jet pointing directly towards the Earth-- are the most numerous among all known gamma-ray sources.

During these years of operations, different point-source catalogs\footnote{All gamma-ray LAT catalogs are publicly available \href{https://fermi.gsfc.nasa.gov/ssc/data/access/lat/}{here}.} have been publicly released, containing from hundreds to thousands of gamma-ray objects, many of them previously unknown \cite{3FGL_paper,Abdollahi_2020}. Subsequent catalogs 
are not mere updates of exposure times, but also the improvement of knowledge about the astrophysical diffuse emission model and better characterization of the instrumental response functions (IRFs). Moreover, specific catalogs may be released, regarding e.g. high-energy sources (\cite{2FHL_paper,3FHL_paper}) or active galactic nuclei (\cite{3lac, 4lac-dr1}).

One of the less known magnitudes in gamma-ray astronomy is distance. As gamma rays do not present any spectral line (an exception being the 511 keV feature \cite{2010arXiv1009.2098S}), the only possible way to determine the distance is by associating the gamma-ray emitter to a known source, which presents absorption or emission lines in other wavelengths, allowing the redshift computation\footnote{In the remaning of the paper, we will use distance and redshift as synonyms. We must keep in mind that the physical distance computation relies not only on the redshift value but on other cosmological parameters such as $H_0$ and $\Omega_\Lambda$.}. Source classes such as BL-Lacs, which do not present emission lines, makes redshift determination very challenging, which lead to the development of photometric estimation techniques, such as photo-z \cite{2018ApJ...859...80K, 2018A&A...611A..97P, 2022arXiv220909877S}.

The knowledge of the redshift is fundamental for studies such as the formation, structure and evolution of early galaxies \cite{2020ApJ...903..128K}, the intergalactic magnetic field \cite{2013MNRAS.432.3485V}, or the Extragalactic Background Light (EBL) \cite{2017MNRAS.470.4089A}, which will be discussed later on this paper.

In recent years, machine learning has become a widely used tool in astrophysics for tasks such as regression, classification, clustering, dimensionality reduction, and feature extraction \cite{2019arXiv190407248B}. Indeed, several well-established algorithms such as K-Nearest Neighbours (KNN), Support Vector Machines (SVMs), Artificial Neural Networks (ANN), Decision Trees (DT) or Random Forests (RF) are widely used for regression problems (see, e.g., \cite{8862451, Sarker_2021} for a review on these techniques). More recent implementations include eXtreme Gradient Boost (\texttt{XGBoost}) \cite{2016arXiv160302754C}, \texttt{LightGBM} \cite{Ke2017LightGBMAH}, and \texttt{CatBoost}\footnote{\url{https://catboost.ai/}} \cite{2017arXiv170609516P}, which allows a natural handling of categorical variables.

Previous works have aimed to predict the redshift of gamma-ray AGNs with machine learning techniques, such as a Superlearner combination of XGBoost, RF, Bayes GLM, and Big LASSO in 4LAC \cite{2021ApJ...920..118D}, Hierarchical Correlation Reconstruction also in 4LAC \cite{2022arXiv220606194D} or SLOPE in 4LAC--DR2 \cite{2022ApJS..259...55N}, with a Pearson correlation coefficient of 0.74 and a RMSE of 0.467.

In a previous paper, we used \texttt{CatBoost}\footnote{CatBoost is an open-source algorithm for gradient boosting decision trees, developed by researchers and engineers at the Yandex technology company \url{https://catboost.ai/}} for a multiclass classification of unIDs \cite{2022MNRAS.515.1807C}. In this paper, we aim to use it for redshift prediction using the latest 4LAC--DR3 catalog \cite{2022arXiv220912070T} (built on the LAT 12-year 4FGL-DR3 source catalog \cite{2022arXiv220111184F}) to train the algorithm with the known-redshift sample of AGNs, and then predict the redshift of the remaining objects. In a second part, we will predict the redshift of the unassociated sample of the 4FGL--DR3 catalog.

The paper is structured as follows: In Section \ref{sec:4lac-dr3}, we describe the 4LAC--DR3 LAT catalog, and select the features to take into account. In Section \ref{sec:catboost_4lac}, we show the \texttt{CatBoost} training, results of the prediction for the remaining 4LAC AGNs, and the implications for the EBL. In Section \ref{sec:catboost_4fgl} the same model, with some modifications, is applied to the unIDs of the 4FGL--DR3 catalog. We conclude in Section \ref{sec:conclusion}.

\section{The 4LAC--DR3 catalog}
\label{sec:4lac-dr3}
The \textit{Fermi}-LAT fourth catalog of AGNs (4LAC) was released in 2019, spanning 8 years of LAT data (2008-2018) \cite{4lac-dr1}. A second release, 4LAC--DR2, was published two years later with two additional years of data \cite{2020arXiv201008406L}, while a third one --and latest at this writing--, 4FGL--DR3, was published in 2022 covering 12 years of LAT observations \cite{2022arXiv220912070T}.

The 4LAC--DR3 contains 3407 individual sources, from which only 1794 present redshift values. In Table \ref{tab:4lac-redshift}, we show by classes the count and percenteage of sources with redshift.

\begin{table}
\centering
\begin{tabular}{ |c|c|c|c|  }
\hline
\rowcolor[gray]{.8} 
\multicolumn{4}{|c|}{Summary of redshift determination in 4LAC--DR3} \\
\multicolumn{1}{|c|}{Class} & Redshift & No redshift & Redshift \%\\
\hline
\rowcolor[gray]{.9} 
agn & 4 & 3 & 57\%\\
bcu & 133 & 1054 & 11\%\\
\rowcolor[gray]{.9} 
bll & 855 & 512 & 62\%\\
css & 5 & 0 & 100\%\\
\rowcolor[gray]{.9} 
fsrq & 749 & 0 & 100\%\\
nlsy1 & 8 & 0 & 100\%\\
\rowcolor[gray]{.9} 
rdg & 37 & 5 & 82\%\\
sey & 1 & 0 & 100\%\\
\rowcolor[gray]{.9} 
ssrq & 2 & 0 & 100\%\\
TOTAL & 1794 & 1574 & 53 \%\\
\hline
\end{tabular}
\caption{Summary of redshift availability in 4LAC--DR3. While many minority classes present a high or perfect degree of completitude, the most abundant classes show scarce redshift values. Overall, almost half of the catalog lacks redshift determination.}
\label{tab:4lac-redshift}
\end{table}

A total of 41 columns are reported, with information such as the detection significance, flux with uncertainty, best-fit spectrum type, best-fit parameters, variability and other spectral parameters. Redundant, version-based or counterpart information columns such as \texttt{'DataRelease', 'RAJ2000'}, etc. are rejected.

Four DR3 new columns, \texttt{['HE\_EPeak', 'Unc\_HE\_EPeak', 'HE\_nuFnuPeak', 'Unc\_HE\_nuFnuPeak']}, give information about the synchrotron peak and flux for curved spectra. Unfortunately, a large fraction ($\sim$15\%) are undetermined, and therefore are removed from the sample\footnote{We performed a test removing this 15\% of null sources and mantaining the variables, but the algorithm didn't yield better results. In fact, they were ranked among the less influential variables (see Section \ref{sec:influence}).}. We use the 24 remaining features as input for the algorithm.

\section{Predicting the redshift of 4LAC sources with \texttt{CatBoost} regressor}
\label{sec:catboost_4lac}

\subsection{Training and validation of the algorithm}
We rely on the \texttt{CatBoost} algorithm \cite{2017arXiv170609516P} to predict the redshift of 4LAC remaining sources. \texttt{CatBoost} is a boosted decision tree machine learning algorithm with native support for GPU executions and categorical labels (e.g., \texttt{SpectrumType}).

First, we mark the categorical variables (\texttt{['SpectrumType', 'Flags', 'CLASS', 'SED\_class']}). We use the sample of redshift-determined AGNs to train the model. A 80/20 train/test split is performed, with a 5--fold cross--validation strategy. Using the \texttt{CatBoostRegressor} class, we perform a grid search varying the \texttt{learning\_rate, depth} and \texttt{l2\_leaf\_reg} hyperparameters.

\begin{figure}
    \centering
    \includegraphics[width=\linewidth]{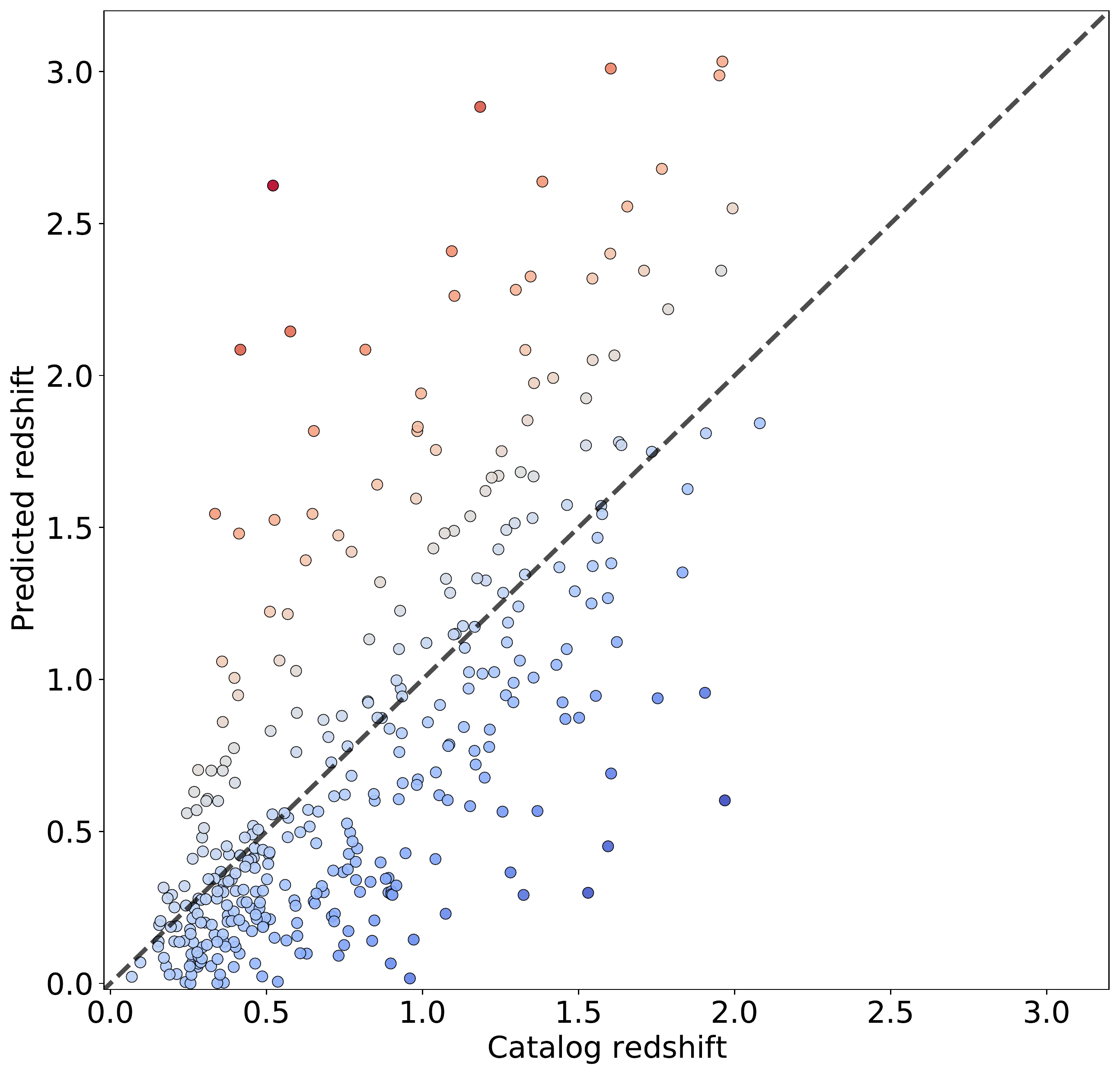}
    \caption{Scatter relation between the \texttt{CatBoost}-predicted (vertical axis) and real (horizontal axis) redshift for the test set of the 4LAC AGN catalog. A diagonal, dashed line represents the equality between the two quantities, i.e., a perfect prediction. Redder colors indicate an overestimation of the redshift, and viceversa for bluer colors.}
    \label{fig:scatter_4lac}
\end{figure}

When predicting on the test set, the algorithm reaches a 0.56 R2 score, with 0.46 RMSE. In Figure \ref{fig:scatter_4lac}, we plot a scatter relation between the predicted and real redshifts for the 4LAC test set. The Pearson correlation coefficient between prediction and actual redshifts is $r=0.71$, which, despite the different approach and newer data release, is a very similar value to those found by \cite{2021ApJ...920..118D} and \cite{2022ApJS..259...55N}. The average redshift value is found to be $z_{avg}=$0.63, with a maximum $z_{max}=$1.97.

We also try a SuperLearner algorithm \cite{polley2010super}, by using an ensemble of eight different \texttt{sklearn} \cite{2012arXiv1201.0490P} regression models: LR, RF, SVR, DT, KNN, ElasticNet, AdaBoost, and Bagging regression. We implement this solution with the \texttt{mlens} package \cite{flennerhag:2017mlens}, and using a LabelEncoder to deal with categorical variables. The results are not as competitive as the \texttt{CatBoost} model, with a R2 score of 0.48 and a RMSE of 0.49.

\subsection{Feature influence and redshift prediction}
\label{sec:influence}
\begin{figure}
    \centering
    \includegraphics[width=\linewidth]{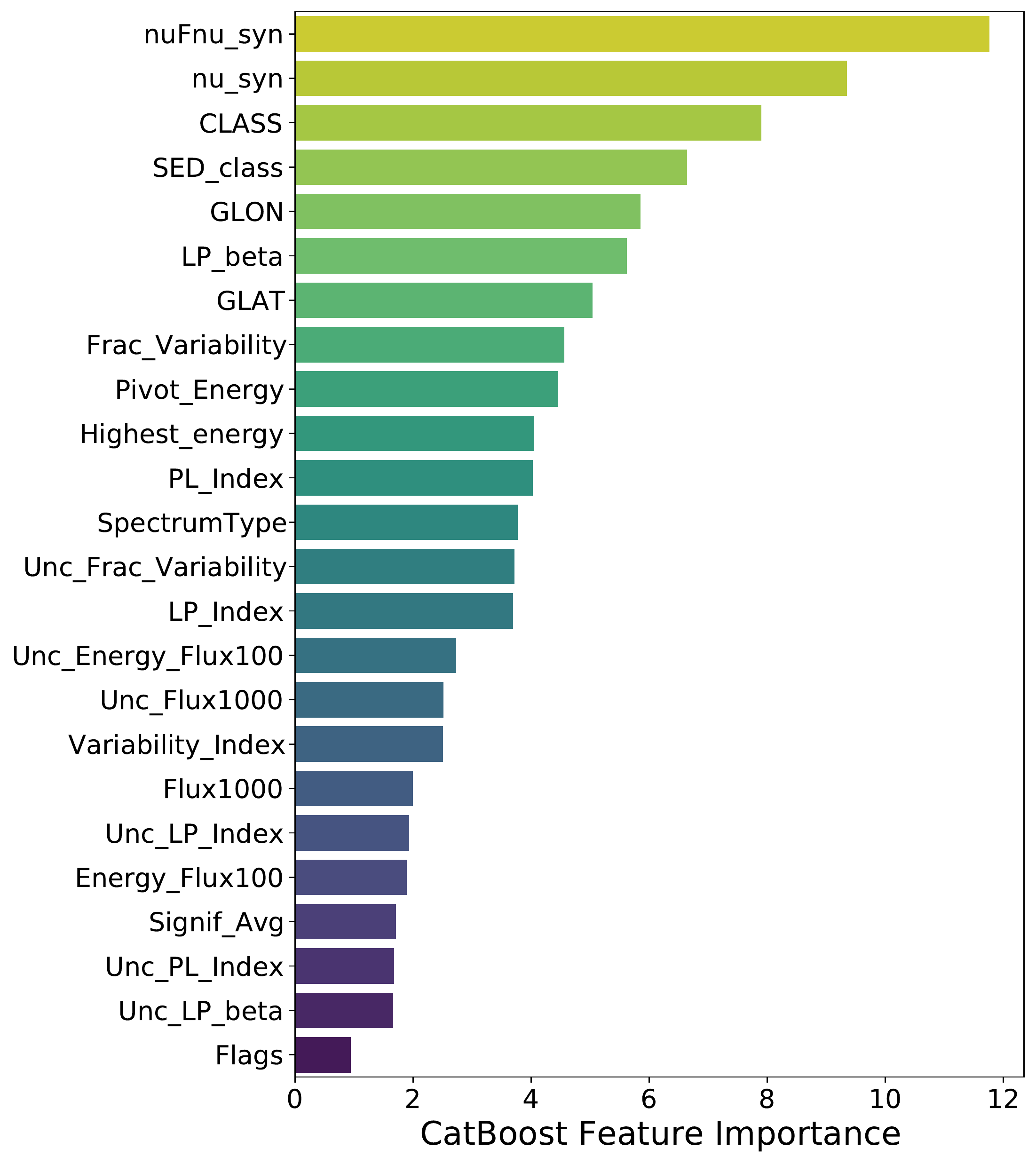}
    \caption{Relative importance of the 4LAC variables in the \texttt{CatBoost} algorithm, in random units.}
    \label{fig:importance_variables}
\end{figure}

By ranking the importance of each feature, we can also see the relative weight of the variables (\texttt{feature\_importances\_} module of the \texttt{CatBoost} regressor) when computing the predicted redshift. The result is shown in Figure \ref{fig:importance_variables}. The most relevant variables turn out to be the ones regarding the class of the source and the synchrotron--peak in the observer frame parameters \texttt{nuFnu\_syn} and {\texttt{nu\_syn}}.

To better understand the importance of the variables when determining the redshift, we use the \texttt{SHAP} package \cite{NIPS2017_8a20a862}. This is an explainer method which quantifies how much is each input feature influencing in the outcome of the \texttt{CatBoost} decision trees.

\begin{figure}
    \centering
    \includegraphics[width=\linewidth]{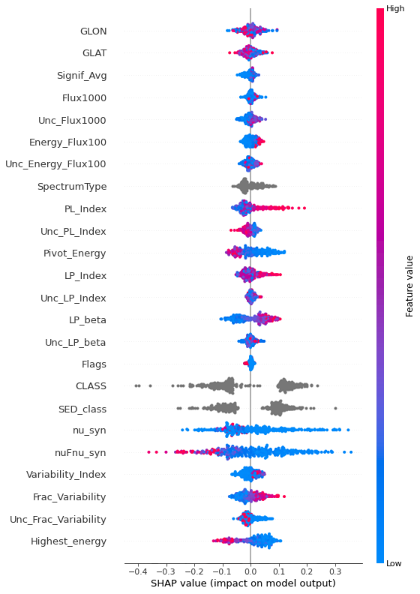}
    \caption{Summary plot of the 4LAC redshift prediction with \texttt{CatBoost}, generated with \texttt{SHAP}. The shown curves are the probability density functions. The gray dots indicate categorical variables, which cannot be quantified as high or low values.}
    \label{fig:shap_summary}
\end{figure}

In Figure \ref{fig:shap_summary}, we show a summary plot of the feature values and their influence on the output redshift. Colors indicate high or low values for each feature (except in the case of categorical variables, where there is no value scale). These values impact on the final result as indicated in the x-axis. The more separated from the zero, the more influential the variable. For example, in the case of \texttt{nu\_syn} (the photon frequency at the synchrotron emission's peak), high values shift the predicted redshift towards lower values, and viceversa. As it is one of the most influential features, the shift can be very large, up to 0.4. In the case of \texttt{Frac\_Variability}, a mid-influential variable, the tend is inverted, but the shift is smaller, up to 0.1.

\begin{figure}
    \centering
    \includegraphics[width=0.9\linewidth]{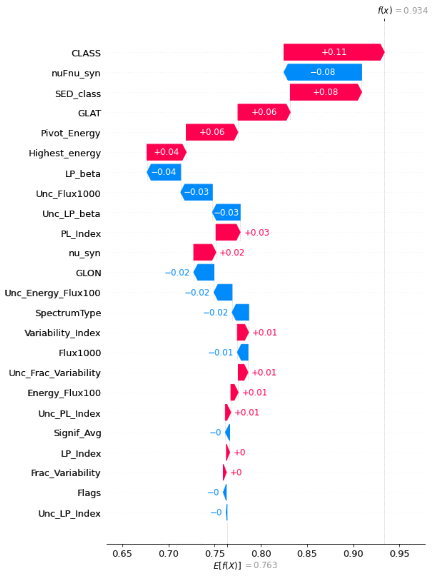}
    \includegraphics[width=0.9\linewidth]{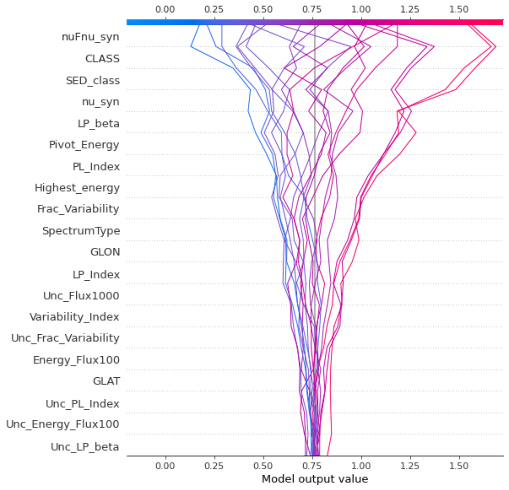}
    \caption{\texttt{SHAP} plots of the 4LAC redshift prediction with \texttt{CatBoost}. \textbf{Upper panel:} value shift for redshift prediction in a random source. Red (blue) color indicate shifts towards larger (smaller) values. \textbf{Lower panel:} redshift value shifts for a random sample of 20 sources.}
    \label{fig:shap_2plots}
\end{figure}

This variable-influence explicability can also be represented for a single source or a small subset of them. In Figure \ref{fig:shap_2plots}, we show two more explainability plots. In the upper panel we show the value shift for each feature. In the lower panel, this is extended to 20 random sources. Starting from a similar value around $z_0\sim0.75$, the sources end up spanning a range $z_{pred}\sim0.2-1.6$. Low influential variables produce small shifts of the redshift value, and viceversa.

\begin{figure}
    \centering
    \includegraphics[width=\linewidth]{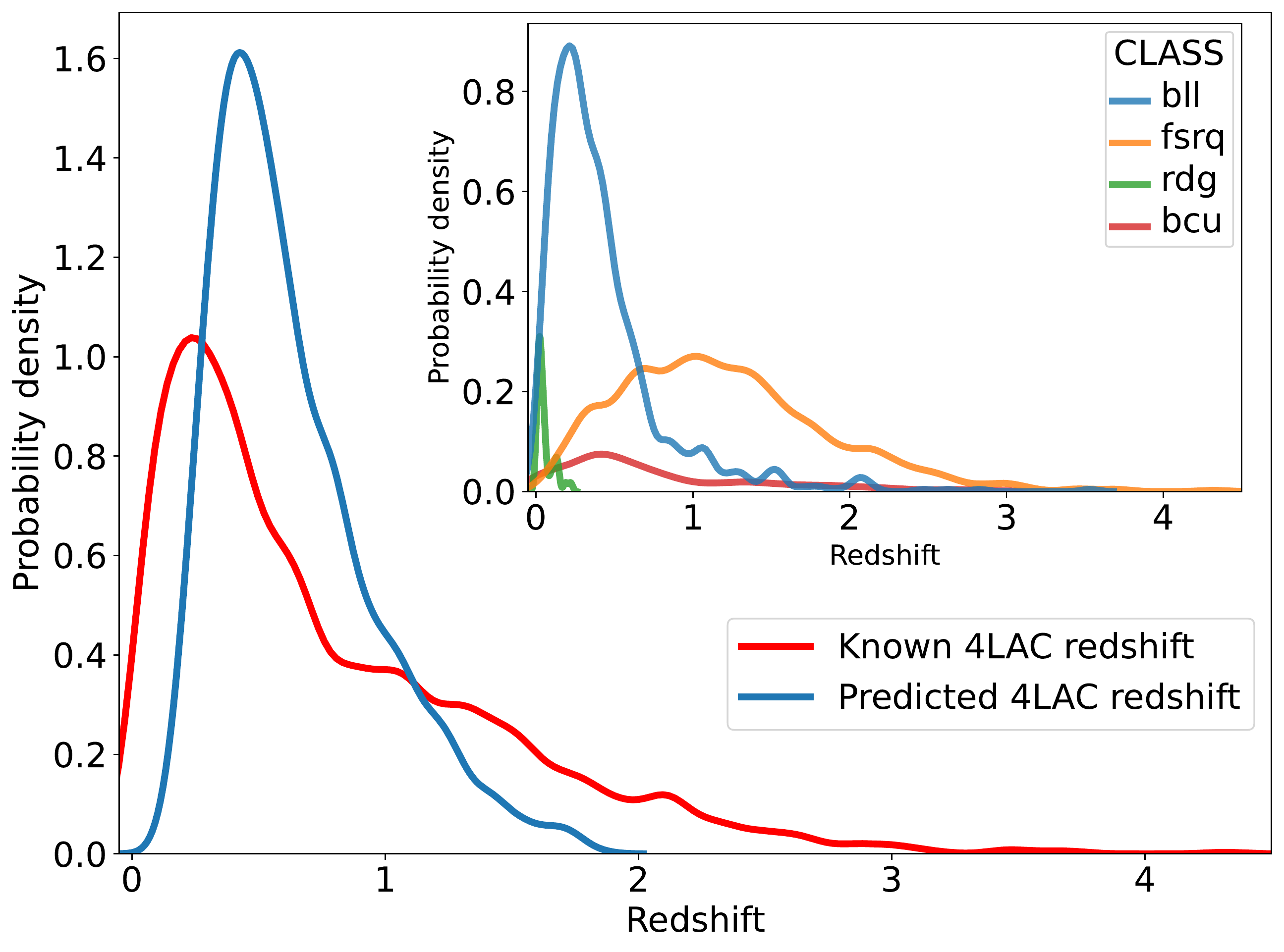}
    \caption{Distribution of the \texttt{CatBoost}-predicted redshift (blue line) and the known redshifts (red line) in the 4LAC catalog. The shown curves are the probability density functions. The subplot shows the distribution of the 4LAC sources with known redshift disaggregated by class.}
    \label{fig:redshift_pred_4lac}
\end{figure}

We now use this model to predict the redshift of the rest of the 4LAC catalog. In Figure \ref{fig:redshift_pred_4lac}, we show the distribution of this predicted redshift together with the known redshifts. The redshift distribution is very different between them, with bll and rdg sources peaking at $z<0.5$, while fsrq and bcu have a more broad distribution which extends up to $z~\sim2.5-3$. This shows the class is a relevant feature for the redshift determination.

The peak of the predictions is shifted towards larger redshifts, yet it does not reach values as high as the known sample. This is most likely because the unknown sample is dominated by bcu, which, if properly classified, would populate other classes in a different proportion that the one observed in the known sample (see the subplot of Figure \ref{fig:redshift_pred_4lac}).

\subsection{Redshift prediction and the Cosmic Gamma-Ray Horizon}
As previously mentioned, one of the main interests of gamma-ray AGN redshift determination is the characterization of the EBL. This is the the cumulative light emitted by stars and accreting compact objects through the whole history of the Universe \cite{1978ApJ...226..609S, 2019PhDT........62D}. These low-energy photons will interact with the gamma rays via inverse Compton scattering, resulting in an effective opacity $\tau$, depending both on the energy and the redshift, which will attenuate the gamma-ray photon flux. Applications of the EBL include the study of star formation \cite{2018Sci...362.1031F} and independent $H_0$ determination \cite{2019ApJ...885..137D}.

In particular, it is interesting to study the so-called cosmic gamma-ray horizon (CGRH), defined as $\tau=1$, i.e., where the probability of a gamma-ray photon to reach us is $1/e$ \cite{2013ApJ...770...77D}. 

The highest energy photons detected for each source as reported in the 4LAC catalog are selected within the the lowest instrumental background photon class (P8R3\_ULTRACLEANVETO\_V2). Comparing this energy for a given source with the energy computed for $\tau=1$ at the same redshift results in a direct test of the
EBL models.

\begin{figure}
    \centering
    \includegraphics[width=\linewidth]{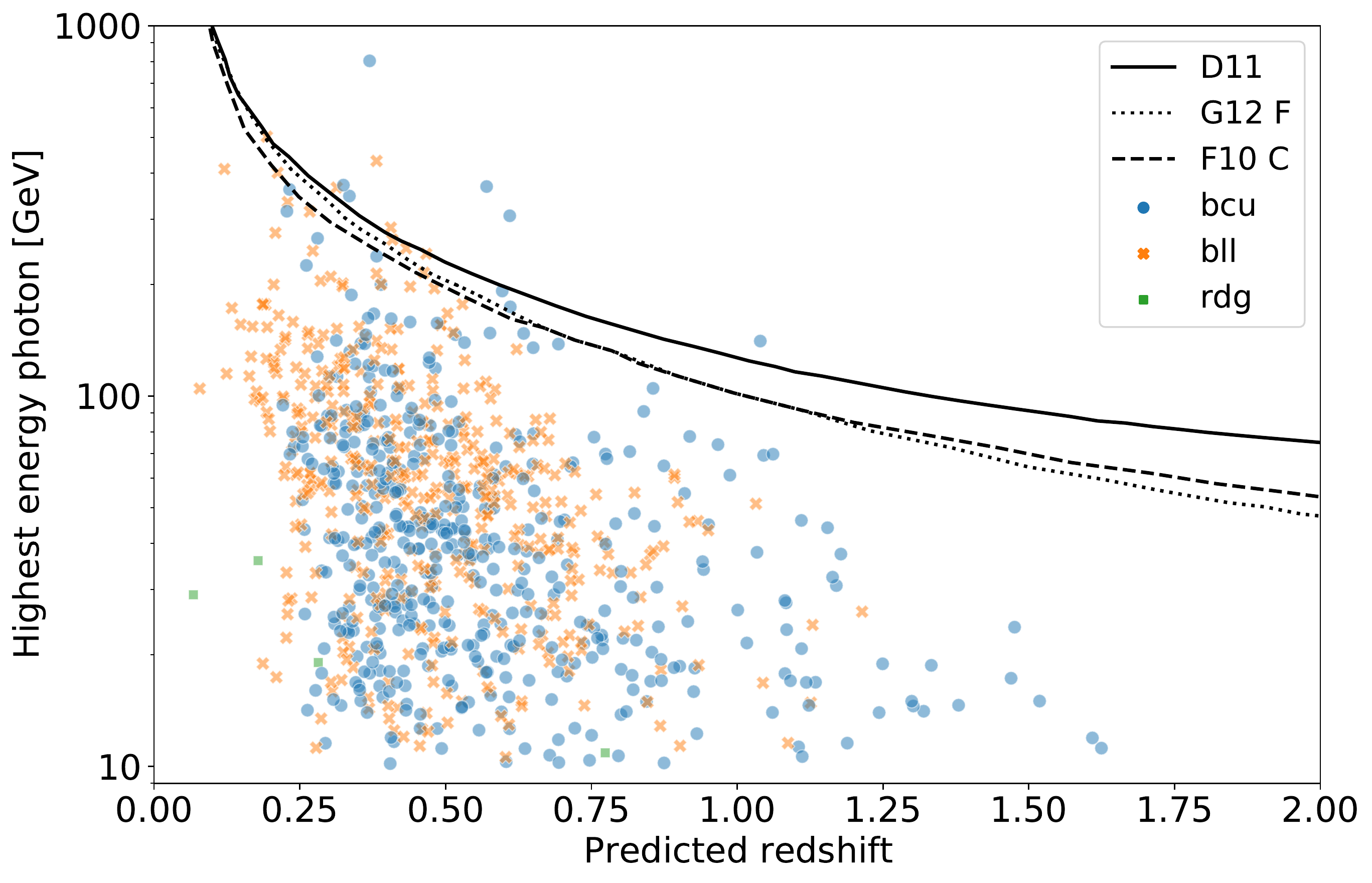}
    \caption{Scatter plot of the highest energy photons for the predicted redshift sample of the 4LAC. Overimposed is the CGRH ($\tau=1$) for three different models, D11, G12, and F10 C (see text for details).}
    \label{fig:ebl_4lac}
\end{figure}

In Figure \ref{fig:ebl_4lac}, we show this plot for the predicted redshift 4LAC sources\footnote{This same plot for the known redshift sources of the 4LAC is presented in \cite{4lac-dr1}.}. Three different EBL models are considered: Dominguez+11 \cite{2011MNRAS.410.2556D}, Gilmore+12 Fiducial \cite{2012MNRAS.422.3189G} and Finke+10 Model C \cite{2010ApJ...712..238F}. Not every 4LAC source has a \texttt{HighestEnergy} value, as only energies larger than 10 GeV are reported; therefore, only a fraction of the sources can be represented. In particular, just three classes appear: bll, bcu and rdg.

As expected, only a handful of sources lie above the CGRH, which, in this case, is not a test for the CGRH itself but rather for the redshift predictions. Should a large fraction of predicted sources exceed the CGRH, it would indicate an statistical overestimation of the redshift.

The one-fold increase of AGNs with redshift estimation would allow a refining of the EBL models and, in the case of applications such as the $H_0$ estimation, would decrease the uncertainty. Such work is out of the scope for this paper.

\section{Predicting the redshift of 4FGL unassociated sources}
\label{sec:catboost_4fgl}
The 4FGL--DR3 \cite{2022arXiv220111184F}, released in 2022, is the most complete compendium of gamma-ray sources. It is composed of 6659 individual sources, from which 2296 (34\%) are unIDs.

In a previous work \cite{2022MNRAS.515.1807C}, we used \texttt{CatBoost} to perform a classification of the 4FGL--DR3 unIDs. As there are 23 different classes present, one of the models was based on a binary unification of AGN-like and PSR-like sources to simplify the casuistic. This allowed the algorithm to reach an average accuracy >99\% when tested on the identified sample.

Therefore, we now use those results to predict the redshift for the AGN-like population, as 4FGL does not report redshift values. We do not predict the redshift for PSR-like sources because LAT detected PSR are Galactic, and their redshift would be negligible. We also do not predict the redshift in this population because PSR-like sources present a highly curved spectrum (best fitted to a power law with super-exponential cutoff model), which the algorithm would not recognize, as it has been trained among AGNs, fitted to a power law or log-parabolic model.

Nevertheless, the \texttt{CatBoost} model must be updated, as 4LAC and 4FGL catalog variables are not exactly the same. Specifically, the following 4LAC variables used in the regression are not present in 4FGL: \texttt{['SED\_class', 'nu\_syn', 'nuFnu\_syn', 'Highest\_energy']}. Additionally, the \texttt{CLASS} variable must be removed, as in the binary classification there is no further distinction rather than AGN-like.

These turned out to be the four most influential variables (see Figure \ref{fig:importance_variables}), and therefore the accuracy of the regression will be hurt. Re-computing the \texttt{CatBoost} model without this variables on the 4LAC sample lowers the R2 score to 0.31, with a RMSE of 0.57. The Pearson correlation coefficient is now 0.55\footnote{We also try the SuperLearner algorithm, which again results in moderately poorer performance.}.

\begin{figure}
    \centering
    \includegraphics[width=\linewidth]{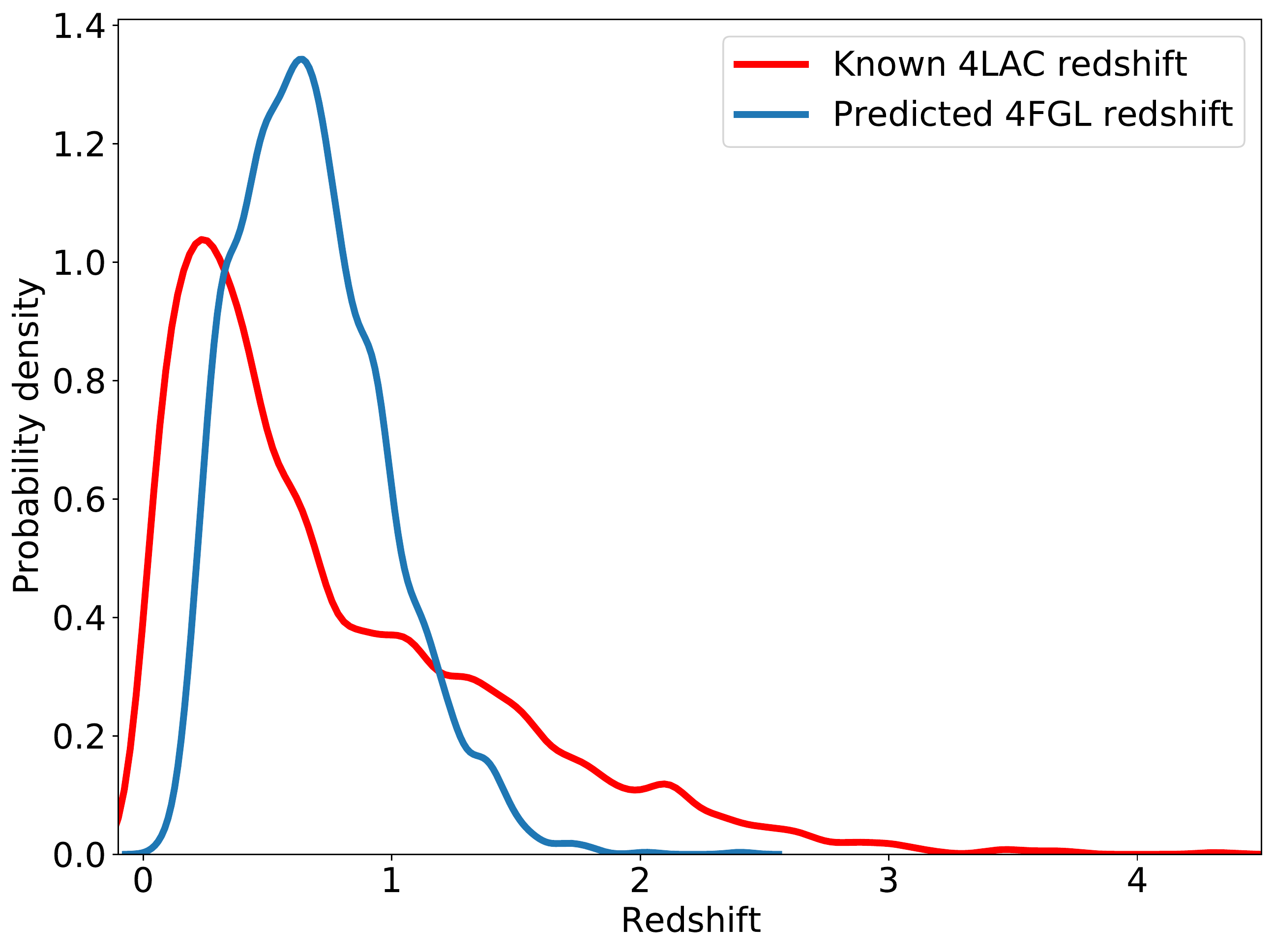}
    \caption{Same as Figure \ref{fig:redshift_pred_4lac}, but for the predictions over 4FGL unIDs.}
    \label{fig:redshift_pred_4fgl}
\end{figure}

We proceed with the redshift prediction for the unID sample of the 4FGL--DR3, keeping in mind the caveats of the poorer performance of the algorithm. In Figure \ref{fig:redshift_pred_4fgl}, we show the redshift distribution of the 4FGL AGN-like unIDs.

The redshift average value is now $z_{avg}=$0.68, with a maximum $z_{max}=$2.40. While the average redshift is very similar to the one found among the 4LAC sample, the maximum value is larger. The overall distribution is similar to the one predicted over the 4LAC, suggesting the underlying population is statistically similar, i.e., the unID AGN-like sample is expected to yield similar redshifts to the 4LAC sample. Indeed, the former, once properly characterized, should be part of the latter.

Exotic sources, such as WIMP dark matter (DM) annihilation or decay \cite{2014arXiv1411.1925C, 2015PNAS..11212264F} could be hidden among the unIDs. In particular, DM subhalos would not allow for redshift determination, as they are devoided of baryonic content and lack any multiwavelength counterpart \cite{2019JCAP...07..020C}. Nevertheless, in principle none of these exotic sources are expected within the AGN-like sample, as DM presents a highly curved spectrum, which would be classified as PSR-like \cite{Mirabal:2013rba}. Such sources can also be studied with ML techniques \cite{2022arXiv220709307G}.

This redshift prediction exercise for the 4FGL catalog unIDs should be seen as a proof of concept, as we are just using the published, condensed information of the LAT sources. If we took the raw, unprocessed data for each 4LAC source, and then applied the resulting algorithm to the same data structure in the case of 4FGL sources, the predictions would be much more precise and reliable.

Additionally, should we have this raw dataset, we would be able to obtain the highest energy photon and reproduce Figure \ref{fig:ebl_4lac} for the full 4FGL unID sample, providing another insight at both redshift predictions and EBL models. Such approach is beyond the scope of this paper and is left for future work.

\section{Conclusions}
\label{sec:conclusion}

In this paper, we have performed a redshift prediction of \textit{Fermi}-LAT gamma-ray sources. The determination of the redshift/distance in gamma-ray astronomy presents a big challenge, due to the necessity of multiwavelength counterpart association and detailed follow-up photometric and/or spectroscopic studies.

In this sense, machine learning algorithms provide a useful tool for redshift estimation, based on the spectral and spatial properties of sources with known redshifts. We used the 4LAC--DR3 catalog, which is the latest \textit{Fermi}-LAT AGN compendium. Half of the catalog lacks redshift values, which can be predicted and compared with the training sample.

To do so, we used \texttt{CatBoost} gradient boosting decision trees, a state-of-the-art algorithm. By selecting 24 features from the 4LAC catalog, we trained the algorithm with the known redshift sample, using a 5-fold cross-validation strategy, a 80/20 train/test split and a grid search for hyperparameter optimization.

When comparing the predictions on the test set, we obtained a R2 score of 0.56 with a 0.46 RMSE. The predicted and real redshifts presented a Pearson correlation of 0.71, similar to previous works \cite{2021ApJ...920..118D, 2022ApJS..259...55N}.

The redshift distribution peaked around $z_{peak}\sim0.5$, with an average value of $z_{avg}=0.63$ and a maximum $z_{max}=1.97$. By using the \texttt{SHAP} explainer package, we were able to understand the most influential variables, and how do they impact on a single prediction depending on the value of each feature.

We also examined the CGRH, defined as the EBL producing a $\tau=1$ photon flux attenuation, with the predicted redshifts. In three different models, only a handful of sources were above the CGRH, which is a robustness test of the predictions. These results could be used to enforce the EBL models and reduce the uncertainties of derived parameters, such as the $H_0$ value or the star formation rate. Such work is out of the scope of this paper.

Finally, we used the results of \cite{2022MNRAS.515.1807C}, where a binary AGN-/PSR-like classification was performed among the unIDs of the 4FGL--DR3, to also predict the redshift of the unID population. We only used AGN-like sources, as the training sample, and reduced the number of variables due to the unmatching between 4LAC and 4FGL catalogs. Unfortunately, the discarded features were precisely the most influential ones, which made the resulting model to present a poorer performance, with a 0.31 R2 score, and 0.57 RMSE, as well as a Pearson correlation coefficient of 0.55. The resulting redshift distribution was statistically similar to that output by 4LAC sources, indicating that the underlying AGN population hidden among the 4FGL unID sample should be expected to also be similar in this sense.

Both the 4LAC and 4FGL predictions --especially the latter-- would certainly improve provided we used the full, raw LAT data instead of a 24/20-variable preprocessed set. This is out of the scope of this paper, and is left for future work.

\section*{Acknowledgements}

The author would like to thank the \textit{Fermi}-LAT Collaboration for the public availability of data. This work made use of NASA’s Astrophysics Data System for bibliographic information.

\section*{Data Availability}

The data used in this article are publicly available at the  \href{https://fermi.gsfc.nasa.gov/ssc/data/access/}{Fermi Science Support Center (FSSC)} of the NASA GSFC.



\bibliographystyle{mnras}
\bibliography{mnras_template} 





\bsp	
\label{lastpage}
\end{document}